%% file: theta.tex
\documentstyle[pre,aps,preprint]{revtex}
\begin{document}
\draft
\input psfig_19

\title{Winding angles for two--dimensional polymers with orientation dependent interactions}
\author{Thomas Prellberg\footnote{email: prel@a13.ph.man.ac.uk} and Barbara Drossel}
\address{Department of Theoretical Physics, University of Manchester, Manchester M13 9PL, UK}
\date{\today}
\maketitle
\begin{abstract}
We study winding angles of oriented polymers with orientation--dependent interaction in two dimensions. Using exact analytical calculations, computer simulations, and phenomenological arguments, we succeed in finding the variance of the winding angle  for most of the phase diagram. Our results suggest that the winding angle distribution is a universal quantity, and that the $\theta$--point is the point where the three phase boundaries between the swollen, the normal collapsed, and the spiral collapsed phase meet. The transition between the normal collapsed phase and the spiral phase is shown to be continuous. 
\end{abstract}

\section{Introduction and summary}
\label{intro}

The study of polymers is one of the most fascinating fields of current
research, because of its relevance not only for material sciences, but
also for the understanding of proteins. Depending on the chemical and
physical environment, a polymer in a dilute solution
can be either swollen or collapsed,
or at the $\theta$--point, which is the boundary point between the
two. Such polymers can be modeled by interacting
self--avoiding random walks (SAWs)
with an interaction energy $\epsilon$ between (non consecutive) bonds that are on the same plaquette of the underlying lattice. As the temperature is decreased, the SAW undergoes the above--mentioned transition at a the $\theta$--temperature, provided that $\epsilon < 0$. 
The value of the
exponent $\nu$ that characterizes the relation between the polymer
length $N$ (monomer number) and its radius of gyration
$R$, is $\nu=$ 3/4, 4/7, and 1/2 above, at, and
below the $\theta$--temperature in two dimensions. 

The phase diagram for the polymer collapse becomes more complex when the polymers are {\it oriented}, 
i.e., when they look different in the two directions along the chain, as, e.g., for A--B polyester \cite{mil91}. In this situation,
the interaction energy between nearby monomers depends  in general on whether their relative orientation is parallel or antiparallel. When the attractive
interaction between parallel monomers is sufficiently strong, the collapsed
polymer winds up to form a spiral. A phase diagram, based on  numerical work and exact results, was suggested in  \cite{ben95}. It contains three phase boundaries, separating the swollen, the normal collapsed, and the spiral phase (the latter being also a collapsed phase), and meeting at one point. 

In that phase diagram, the line along which parallel and antiparallel interactions are equally strong plays no special role. If this is correct, the values of the critical exponents do not fall into universality classes that are determined by symmetries. In fact, conformal field theory \cite{car94} suggests that the 
exponent associated with the partition function (usually denoted $\gamma$) may depend continuously on the parallel interaction energy in the swollen phase, 
while the exponent $\nu$ remains constant. This supposed nonuniversality of $\gamma$ in the swollen phase is complemented by the observation that at the collapse transition $\gamma$ assumes different values on different lattices: Its value on the square lattice is different from that
on the so--called Manhattan lattice, where each bond of the lattice has a preassigned orientation, thus naturally excluding parallel contacts. Numerical studies of the swollen phase \cite{ben95,fle95,bar96,tro97}, however, show no indication of a variation of $\gamma$ with the parallel interaction strength.  These studies involve exact enumeration \cite{ben95}, Monte Carlo simulation \cite{fle95,bar96}, and transfer matrix calculation on a strip of finite width \cite{tro97}. 
In all these studies, as well as in our own exact enumerations, the (very small) variation of $\gamma$ with the parallel interaction strength decreases with increasing polymer length, making it unlikely that $\gamma$ should show nonuniversal behavior for much larger polymer length. 

Furthermore, transfer matrix calculations on the collapse line \cite{tro97} suggest that, whenever the antiparallel interactions energy is lower than the parallel one, $\gamma$ has the same value as on the Manhattan lattice, and that the $\theta$--point (where parallel and antiparallel interactions are equally strong) is the point in the phase diagram where the three phase boundaries meet. If this scenario is correct, the exponent $\gamma$ is a universal quantity, and its value at the collapse transition depends only on whether the symmetry between parallel and antiparallel interactions is broken. 

In this paper, we study
the {\it winding angle distribution} for polymers  in two dimensions with
orientation--dependent short--range interactions. Two monomers that have
a parallel contact are connected by a loop that encloses one of the end points of the polymer, i.e. their winding angle differs by $2\pi$. 
Since the winding angle is so closely related to the occurrence of parallel contacts, it should be equally sensitive to a change in the parallel interaction strength as the exponent $\gamma$. In fact, we find analytically a different winding angle distribution at the collapse transition on the Manhattan lattice and on the square lattice. As for the exponent $\gamma$, the question arises whether the winding angle distribution is determined by simple universality criteria. 
Usually, winding angle
distributions depend only on universal features like symmetries and interaction range, when the length of the
polymer is sufficiently large \cite{dro96}.  The important result of this paper is that the winding angle distributions for oriented interacting polymers are also universal. The main evidence comes from the collapsed phase. Minimizing the free energy, we find that 
the winding angle distribution in the collapsed phase depends only on whether the symmetry between parallel and antiparallel contact energies is broken. 
Just as the transfer matrix calculation of \cite{tro97}, this  modifies
the phase diagram suggested  in  \cite{ben95} and moves the phase boundary between the normal collapsed and the spiral phase on the symmetry line, where parallel and antiparallel interactions are equally strong. 
Our argument shows also that the phase transition along this line is continuous, in contrast to \cite{ben95,tro97}. The $\theta$--point is the point where the three
phase boundaries meet. Due to its special role, it can have an exponent $\gamma$ and a winding angle distribution that is different from all other points, without leading to nonuniversal behavior. 

Certainly, this phase diagram is very appealing due to its simplicity. Remarkably, we arrived at our results independently of \cite{tro97}, and prior to learning about that work. While \cite{tro97} provides the stronger numerical evidence for the special role of the $\theta$--point, our work gives insights in the underlying physics of oriented polymers:
Whenever the antiparallel interaction energy is more negative than the parallel one, the winding angle of the polymer decreases during the collapse. It is confined in the collapsed phase and becomes zero in the ground state, where the end points must be at the surface. When the two interactions are equal, however, the variance of the winding angle increases during the collapse, and is always proportional to $\ln N$. When the parallel interaction is stronger than the antiparallel one, the collapsed phase has an overall spiral shape, the number of parallel contacts being proportional to the length of the polymer. 

The outline of this paper is as follows: In section \ref{definitions}, we describe some basic properties of self--avoiding interacting walks, introducing and discussing the partition function and the general form of the winding angle distribution.
In section \ref{LSAT}, we derive an exact expression for the winding--angle distribution at the collapse transition in the absence of parallel contacts between monomers. We also show results of a Monte Carlo simulation that agree well with this analytical result. 
In section \ref{arguments}, we conjecture the variance of the winding angle
 for most  of the 
phase diagram, and we argue that the transition from the normal collapsed to the 
spiral state occurs when parallel and antiparallel interactions become equally strong. We compare 
to the phase diagram obtained from exact enumeration and explain the origin of the discrepancies. Section \ref{conclusion} summarizes and discusses our results.

\section{General properties of interacting self--avoiding walks}
\label{definitions}
\subsection{The partition function and the critical exponent $\gamma$}
\label{partition}

\begin{figure}
\centerline{\psfig{file=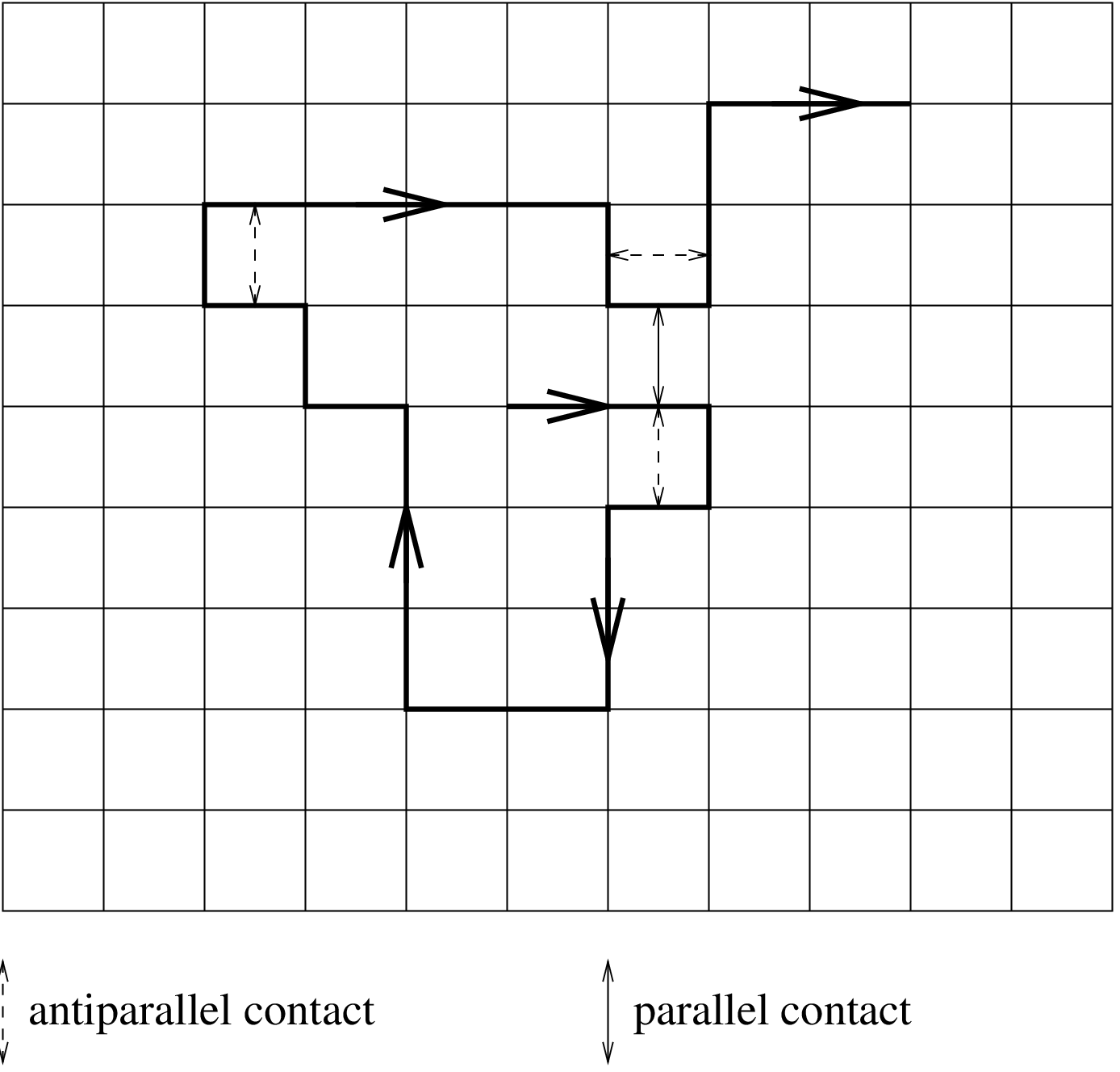,height=3.7in,angle=-90}}
\caption{An oriented self--avoiding path on a square lattice}
\label{path}
\end{figure}

We model a polymer in two dimensions by a SAW on a two--dimensional square lattice, as shown in Fig.~\ref{path}. (In some cases, also a hexagonal lattice is chosen.)
The steps of the walk, which coincide with lattice bonds, can be viewed as monomers. Each site of the lattice can only be visited once. This condition models 
the excluded volume effect of the polymer. Starting at one end point and stepping along the trajectory of the SAW, each bond can be assigned the direction 
in which it is passed. By this procedure, the polymer obtains an orientation. Short--range interactions between monomers are taken into account by assigning an energy $\epsilon_a$ or $\epsilon_p$ to each pair of (non consecutive) monomers that lie on the same plaquette and that have an antiparallel or parallel relative orientation. The weight for each parallel (antiparallel) contact is
 given by $\omega_p=\exp(-\epsilon_p/k_BT)$ 
($\omega_a=\exp(-\epsilon_a/k_BT)$), where $T$ is the temperature and $k_B$ is
the Boltzmann constant.

The partition function for a polymer of $N$ steps is then 
\begin{equation}
Z_N = \sum_{m_a, m_p} g_N(m_a,m_p)\omega_a^{m_a} \omega_p^{m_p}\, ,
\end{equation} 
where $m_a$ and $m_p$ are the number of antiparallel and parallel contacts,
and $g_N(m_a,m_p)$ is the number of configurations (starting at a given point)
with these contact numbers. 

For $\omega_a=\omega_p=0$, one has a normal SAW without any interaction except the self--avoidance, and the partition function is identical to the total 
number of SAWs starting at a given point, which is known to be
\begin{equation}
Z_N \approx A \mu^N N^{\gamma-1}\, , \label{gamma}
\end{equation}
with $A\approx 1.771$, $\mu \approx 2.638$ \cite{ent93}, and $\gamma = 43/32$ \cite{nie82}. This form of the partition
function is believed to hold also for $\omega_a,\omega_p \neq 1$, as long as the polymer is in the swollen phase, with different values of $A$ and $\mu$.

The mean number of antiparallel and parallel contacts in the swollen phase is 
\begin{eqnarray}
\langle m_i\rangle &=& \omega_i {\partial \ln Z_N \over \partial \omega_i}\nonumber\\
&\approx & \omega_i \left({\partial \ln A \over \partial \omega_i} + N {\partial \ln \mu \over \partial \omega_i} + \ln N {\partial \gamma \over \partial \omega_i}\right)\,, \label{meancontacts} 
\end{eqnarray}
with $i=a$ or $i=p$. 
While $A$ can depend on both $\omega_p$ and $\omega_a$, the free energy per monomer $\mu$ does not change with $\omega_p$. This was exactly proven in \cite{ben95} for $\omega_p \le \omega_a$ with $\omega_a = 1$, and --- assuming the scaling form Eq.~(\ref{gamma}) --- it implies that the number of parallel contacts increases not faster than logarithmically with $N$. 
This is not surprising when one realizes that a SAW can have antiparallel contacts anywhere along its trajectory (i.e., $\langle m_a\rangle \propto N$), while the average number of parallel contacts should not increase faster than the number of windings, which in turn increases logarithmically in $N$. 

On a more phenomenological level, one can make the following entropy consideration: as long as 
$\omega_p \le \omega_a$, there is no energetic disadvantage for the polymer 
to have its endpoints at the surface, in which case it has no parallel
contacts. Such configurations correspond to the case $\omega_p=0$. We can therefore assume that the probability that a randomly chosen configuration (for $\omega_a \ge \omega_p > 0$) 
has both end points at the surface is not smaller than 
the square of the ratio of the surface area to the volume, proportional to
$R^{-2} \propto N^{-2\nu}$. The entropy loss per monomer due to the restriction
of the endpoints to the surface is of the order $2\nu\ln N /N$ and vanishes in the thermodynamic limit $N \to \infty$, and so does the change in free  energy per monomer. This argument is not restricted to the swollen phase. 
Consequently,  the phase boundary between the swollen and the collapsed phase cannot depend on $\omega_p$. 

\begin{figure}
\centerline{\psfig{file=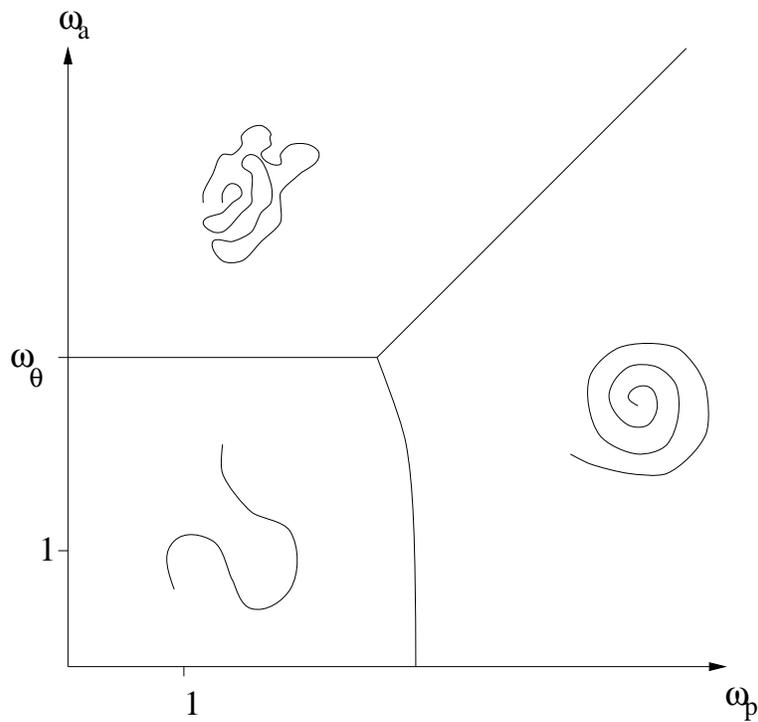,height=3.7in,angle=-90}}
\caption{Qualitative phase diagram of interacting self--avoiding walks}
\label{phasediagram}
\end{figure}

When $\omega_p$ is increased beyond $\omega_a$, there must be a point where the free energy becomes dependent on $\omega_p$, as proved in \cite{ben95}. This point, which is a nonanalyticity of the free energy,  marks the phase transition to the spiral phase (if one assumes the simplest scenario of only one phase transition along a line $\omega_a = $const).  
The qualitative phase diagram is shown in Fig.~\ref{phasediagram}. 

So far, we have not yet discussed the last term in Eq.~(\ref{meancontacts}). It seems implausible that a critical exponent, which is a universal quantity,
should vary within one phase, and this term should therefore vanish. In particular, a repulsive interaction between monomers (i.e., $\omega_{p,a} < 1$) is nothing else than an increased excluded
volume, which can hardly modify the value of a critical exponent. If $\partial \gamma / \partial \omega_p$ vanishes, the mean number of parallel contacts is independent of $N$ for large $N$, i.e., it saturates, in agreement with the above--mentionend numerical results \cite{ben95,fle95,bar96,tro97}. If $\partial \gamma / \partial \omega_p$ did not vanish, as suggested by the conformal field theory \cite{car94}, the number of parallel contacts would increase logarithmically in $N$. An increase in the excluded volume for {\it parallel} contacts, i.e., an increase in $\omega_p$ by $\delta \omega_p$,
 would then have a similar effect as an increase in polymer length from  $N$ to $N + a \ln N$ in the expression for $Z_N$ above, 
with $a=(\partial \gamma / \partial \omega_p)(\delta\omega_p/\ln \mu)$.

The following argument suggests that the number of parallel contacts saturates in the limit $N \to \infty$. 
 We assume that the polymer is radially scale invariant, i.e. that it is statistically mapped onto a polymer of length $bN$, when the coordinates $\vec r$ of all monomers are scaled to $b^\nu \vec r$ (the starting point of the polymer being the origin of the coordinate system). This is equivalent to the statement that, when represented in the  $\theta - \ln r$--plane, the polymer is translation invariant in $\ln r$--direction.
For each value of $r$, there exists consequently the same mean number of parallel ``close encounters'' over a given distance $\Delta (\ln r)$. However, only for $\Delta (\ln r) = 1/r$ such a close encounter corresponds to a parallel contact.
The total number of parallel contacts is then proportional to 
$\int_0^{\nu\ln N}(1/r)d(\ln r) $ and converges for $N \to \infty$. This argument is in agreement with the numerical observation in \cite{bar96} that the total number of ``loops'' saturates for large $N$. 
We conclude that the universality hypothesis holds also for the exponent $\gamma$, and that $\gamma$ is constant in the swollen phase, provided that our assumption of radial scale invariance is correct.
It might be that the conformal field theory in  \cite{car94} does not capture correctly the difference between parallel contacts and ``close encounters''.

At the $\theta$--point $\omega_p=\omega_a\equiv \omega_\theta$, the value of $\gamma$ (denoted $\gamma_t$) is known to be $\gamma_t = 8/7$ \cite{dup87}. It is also known that $\gamma_t=6/7$  for a self--avoiding walk that is part of a percolation cluster hull and therefore has no parallel contacts \cite{bra89}. 
Since $\nu = 4/7$ \cite{sal87} for these walks, they are at the collapse transition. 
The authors of \cite{tro97} argue that $\gamma_t = 6/7$ for all $\omega_p < \omega_a = \omega_\theta$. If this is correct, the partition function contains a 
crossover term close to the $\theta$--point $\omega_p = \omega_a = \omega_\theta$,
\begin{equation}
Z_N \approx  \mu^N N^{1/7}f(\Delta \omega N^\Psi)\,, \label{crossover}
\end{equation}
where $f(x)\approx const.$ for small $x$, and $f(x) \propto x^{-2/7\Psi}$
for large $x$, and $\Delta \omega = \omega_a - \omega_p$. $\Psi$ is a 
crossover exponent.
A calculation analogous to Eq.~(\ref{meancontacts}) gives then
$\langle m_p \rangle \propto N^\Psi$ at the $\theta$--point. 
Since the mean number of parallel contacts saturates in the swollen phase and 
 increases as $\sqrt N$ in the collapsed phase (see below), we expect $0< \Psi < 1/2$. For a related problem, the adsorption of a self--interacting polymer at a surface, the corresponding exponent has the value $\Psi=8/21$ \cite{van91}. Since the adsorption of a polymer at its own surface (i.e., spiral formation),
is somewhat different, the two crossover exponents need not be the same.

In the low--temperature phase, the polymer has a finite density. Therefore
$\nu = 1/2$, and surface effects become important. The partition function 
is assumed to have the general form \cite{owc93}
\begin{equation}
Z_N \approx A \mu^N \kappa^{\sqrt{N}} N^{\gamma_- -1}\, . \label{gamma2}
\end{equation}
We will argue below that for $\omega_p=\omega_a$ the number of parallel contacts is proportional to $\sqrt{N}$, while it saturates for $\omega_p < \omega_a$. 

\subsection{Winding angle distributions}
\label{windings}

For self--avoiding walks, the winding angle distribution is generally
described by a Gaussian
\begin{equation}
p(\theta) \propto \exp(-\theta^2/2 C \ln N)\, , \label{Gauss}
\end{equation}
with a variance $C \ln N$ \cite{fis84,dup88}. Such Gaussian distributions with a variance proportional to $\ln N$ occur  generically for radially  scale invariant polymers, given that the winding center is visited only a finite number of times \cite{dro96}. The value of $C$ is $C=2$ in the swollen phase and $C = 24/7$ at the $\theta$--point of a polymer with no orientation
dependence in the interaction \cite{dup88}, as obtained from an analytical calculation on an hexagonal lattice. 
The value $C=2$ is nicely confirmed by an exact enumeration on a square lattice, where we obtain $C = 2.0005(6)$ for polymers up to length $N=26$, using differential approximant analysis \cite{gut89}. 

The value of $C$ in the collapsed phase (for $\omega_p=\omega_a$) is not known, but it is larger than the previous two values, since the winding angle apparently increases during the collapse. For dense SAWs, $C$ is known to be $C=4$ \cite{dup88}. Since they have no self--interaction apart from self--avoidance, dense polymers have their finite density due to an external pressure and are ``hot'', in contrast to collapsed polymers, where the attraction between monomers determines the density. For this reason, the density of monomer--monomer contacts is different in both cases, and it is not clear whether $C$ can be the same. 

In the next section, we will derive $C=6/7$ for a self--avoiding walk that is part of a percolation cluster hull and therefore has no parallel contacts. Since $\nu = 4/7$ for these walks, they are at the collapse transition. 
The value of $C$ in most other parts of the phase diagram will be discussed in Sec.~\ref{arguments}.

\section{The winding angle distribution at the collapse transition}
\label{LSAT}

\begin{figure}
\centerline{\psfig{file=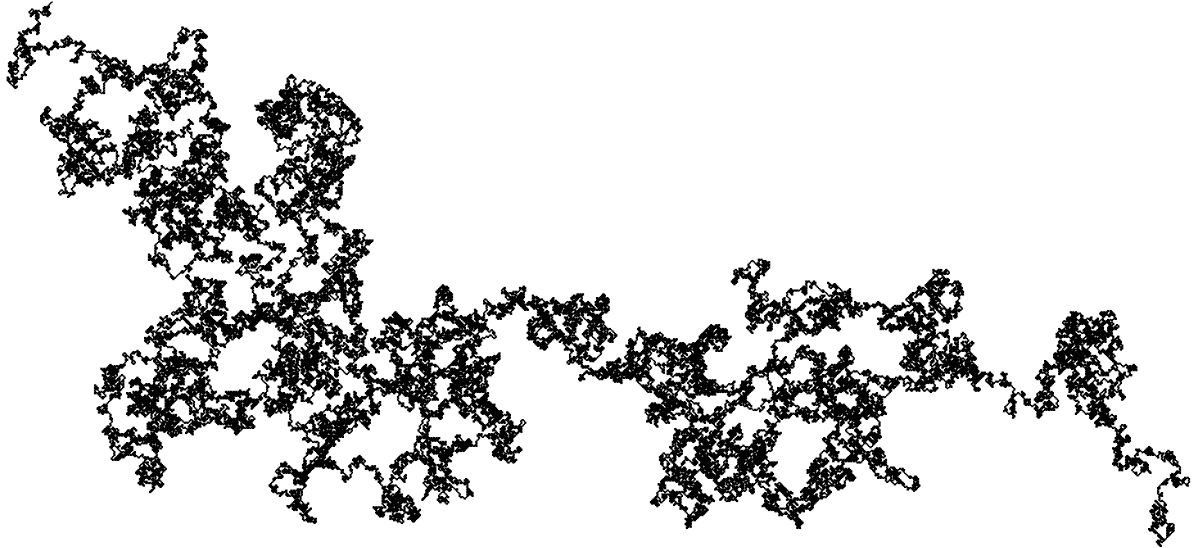,height=4.2in,angle=-90}}
\caption{An interacting self--avoiding walk on the Manhattan lattice at the collapse transition}
\label{manhattan}
\end{figure}

In this section, we calculate the winding--angle distribution for a SAW on a Manhattan lattice at the collapse point. Fig.~\ref{manhattan} shows such a walk. The Manhattan lattice is an array of alternating one--way streets, thus not allowing parallel contacts and always keeping the way back to the origin open. 
SAWs on this lattice can be grown kinetically in a very efficient way, since they get trapped only through loop formation. One starts at the origin and constructs a path by going at each step in one of the allowed directions. When the path closes to a loop, it is cancelled. Since this procedure gives contacts a higher statistical weight than free steps, the polymer has an effective interaction. In fact, one can show that the path can be mapped onto the perimeter of a percolation cluster \cite{bra89}, which in turn is known to have an exponent $\nu=4/7$ \cite{sal87}. 
This means that the polymer is at the collapse point. 

\begin{figure}
\centerline{\psfig{file=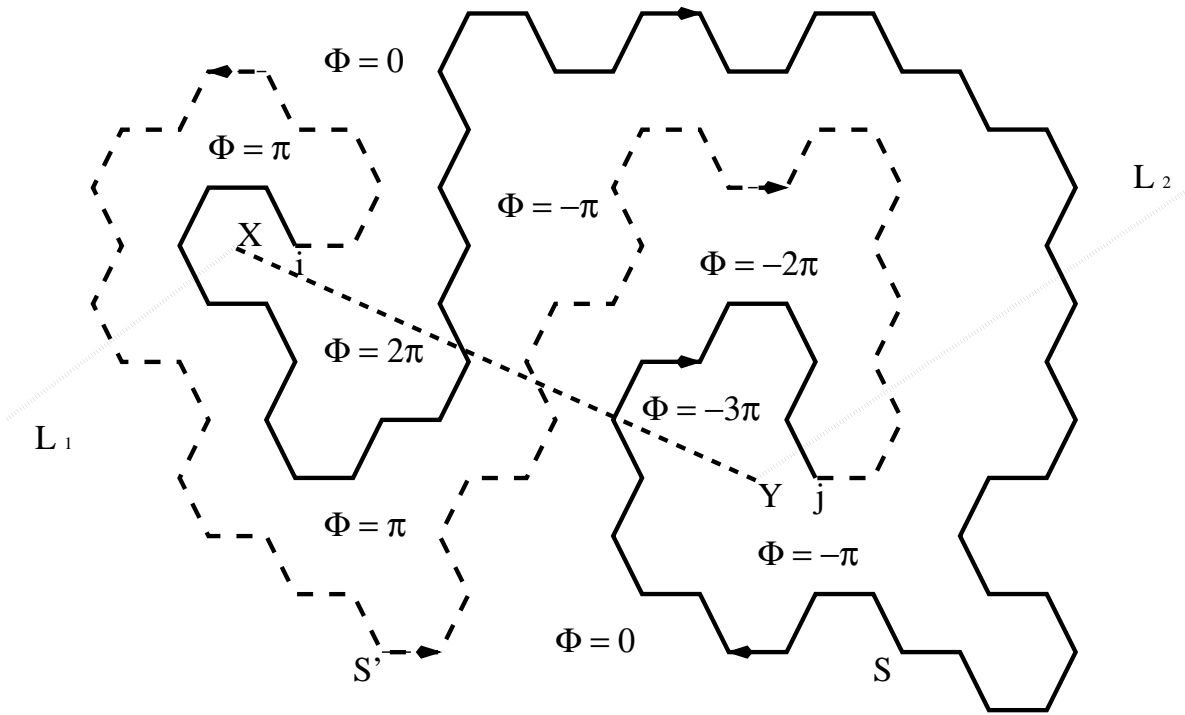,height=3.7in,angle=-90}}
\caption{Illustration of the calculation of the winding angle distribution}
\label{hexa}
\end{figure}

In order to find its winding angle distribution, we have to calculate the winding angle distribution of parts of percolation cluster hulls. The procedure is similar to the one in \cite{dup88,sal87}, and is conveniently performed on an hexagonal lattice. We start with the $O(n)$--loop model \cite{nin84}
with the partition function 
\begin{eqnarray}
Z_{O(n)} &=& \int\prod_kd^NS_k\prod_{\langle i,j \rangle} (1+\beta \vec S_i \cdot \vec S_j)\nonumber \\
&=& \sum_{\cal G}\beta^l n^P\label{Z}\,.
\end{eqnarray}
$i,j,k$ are lattice sites, $\langle i,j \rangle$ nearest neighbors, and $\vec S$ an $n$-vector: $|\vec S|^2=n$. In the second line, the sum is performed over all graphs ${\cal G}$ of $P$ non--intersecting polygons of total length $l$.
For $n\in [-2,2]$, the loop model has a critical point $\beta_c=[2+(2-n)^{1/2}]^{-1/2}$. 
For $n=1$ and $\beta=1$, each loop has the same weight. These loops can be interpreted as percolation cluster hulls for site percolation on the dual triangular lattice at the percolation threshold $p_c=1/2$. We are interested in the winding--angle distribution of a segment of a loop. Fig.~\ref{hexa} illustrates 
the following calculation. We look at the function
\begin{equation}
G_{O(n)}(\vec X - \vec Y,e_1,e_2) = \sum_{{\cal G}_1}
 W_{O(N)}({\cal G}_1) \exp[ie_1\pi (n_1+n_1') + i e_2 \pi (n_2+n_2')]\, \label{corr}
\end{equation}
${\cal G}_1$ are like the graphs  ${\cal G}$ above, but with both points $i$ and $j$ lying on the same loop. Both parts ${S}$ and ${S'}$ of the loop are given the same orientation from $i$ to $j$. $n_1$ ($n_1'$)and $n_2$ ($n_2'$) count the number of intersections of the oriented paths ${S}$ (${S'}$) with ${L}_1$ and ${L}_2$ respectively, crossing in different directions having a different sign. Without the phase factor, $G_{O(n)}$ is a four--spin correlation function, with two spins at $i$ and two spins at $j$. 

To calculate Eq.~(\ref{corr}), we transform it into a solid--on--solid (SOS) model. Height variables $\phi$ are defined on the centers of the hexagons, such that two adjacent heights are equal or different by $\pm \pi$. The polygons, once arbitrarily oriented, become domain walls with a step of $+\pi$ on the left of any oriented line. Along the straight line connecting $\vec X$ to $\vec Y$, the height changes by $2\pi$. In the SOS language, this corresponds to a dislocation line with a vortex at $\vec X $ and an antivortex at $\vec Y$ \cite{sal87}. At the vertices of the honeycomb lattice, the SOS walls turn by $\pm \pi/3$. The SOS weight $W_{\text{SOS}} $ is calculated as the product along the walls of local factors $\beta \exp(iu)$ ($\beta \exp(-iu)$) at each left (right) turning vertex. Summing over the two independent orientations of each polygon (except, of course, the special polygon connecting $i$ and $j$) gives a phase factor
 $2\cos 6u$ for each polygon. The SOS weight of 
a graph ${\cal G}_1$ is then
$$
W_{\text{SOS}}({\cal G}_1) = \beta^ln^P \exp[iu(n_+-n_-) + iu(n_+'-n_-') + 6iu(P_+ - P_-)]\,.
$$
$n_+$ and $n_-$  ($n_+'$ and $n_-'$) are the total number of local left and 
right turns of path ${S}$ (${S'}$).
 $P_+$ and $P_-$ are the total number of right and left  polygons surrounding $i$ and $j$. In the asymptotic limit, we have $n_+ - n_-=6(n_1+n_2$) and 
 $n_+' - n_-'=6(n_1'+n_2'$).  
Let us define the SOS correlator 
$$G_{\text{SOS}}(\vec X - \vec Y,e_1', e_2') = \sum_{{\cal G}_1}
 W_{\text{SOS}}({\cal G}_1) \exp(ie_1'\phi(\vec X) + i e_2' \phi(\vec Y) )\,.
$$
Equating this correlator with Eq.~(\ref{corr}) above gives
$n=2\cos 6u$ and $e_1'=e_1+e_0$,  $e_2'=e_2+e_0$, $e_1+e_2=0$. 
The new constant $e_0$ is $e_0=-6u/\pi$. 

Now, at $\beta = 1$, the SOS model renormalizes onto the low--temperature phase of the Coulomb--gas model for $g = 2/3$ (if $n=1$) \cite{nin84}, and
\begin{equation}
G_{\text{SOS}}(e_1',e_2') = |\vec X - \vec Y|^{e_1'e_2'/g+gm_1m_2}\,.\label{G}
\end{equation}
The magnetic charges $m_1$ and $m_2$ are $m_1=-m_1=1/2$ \cite{sal87}, 
due to the vortex pair. 

The winding angle is finally extracted from 
$$
\langle \exp[ie\pi(n_1-n_2 + n_1'-n_2')]\rangle_{O(N)}
= \exp(-e^2g^{-1}\ln|\vec X - \vec Y|)\,.
$$
Fourier transformation yields immediately a Gaussian distribution for $n_1-n_2+n_1'-n_2'$.
Each of the two paths ${S}$ and ${S'}$ has the same number of intersections with ${L}$ and ${L'}$, and therefore $n_1 = n_1'$ and $n_2 = n_2'$. (In certain situations, $n_1$ and $n_1'$ differ by one, but this effect can be neglected in the thermodynamic limit, where the number of intersections becomes very large.)
 In terms of the angle $\hat \theta = \pi(n_1-n_2+n_1'-n_2')$ the resulting distribution reads
$$
P(\hat\theta) = (16\pi g^{-1} \ln|\vec X - \vec Y|)^{-1/2} \exp(-g \hat \theta^2/16 \ln|\vec X - \vec Y|)\,.
$$
Since both paths make exactly the same contribution to $\hat \theta$, the winding angle of one path is given by $\hat \theta/2$. 
For large distances $|\vec X - \vec Y|$, the windings around $\vec X$ and $\vec Y$ are independent from each other and have the same probability distribution. Replacing  $|\vec X - \vec Y|$ by $N^\nu$ ($N$ being the length of the polymer), we arrive at the winding angle distribution of path ${S}$ around point  $\vec X $,
\begin{equation}
P(\theta) = (4\pi \nu g^{-1} \ln N)^{-1/2} \exp(-g\theta^2/2\nu \ln N)\,.
\end{equation}
Inserting $g=2/3$ and $\nu = 4/7$, we find $C = 6/7$. 

\begin{figure}
\centerline{\psfig{file=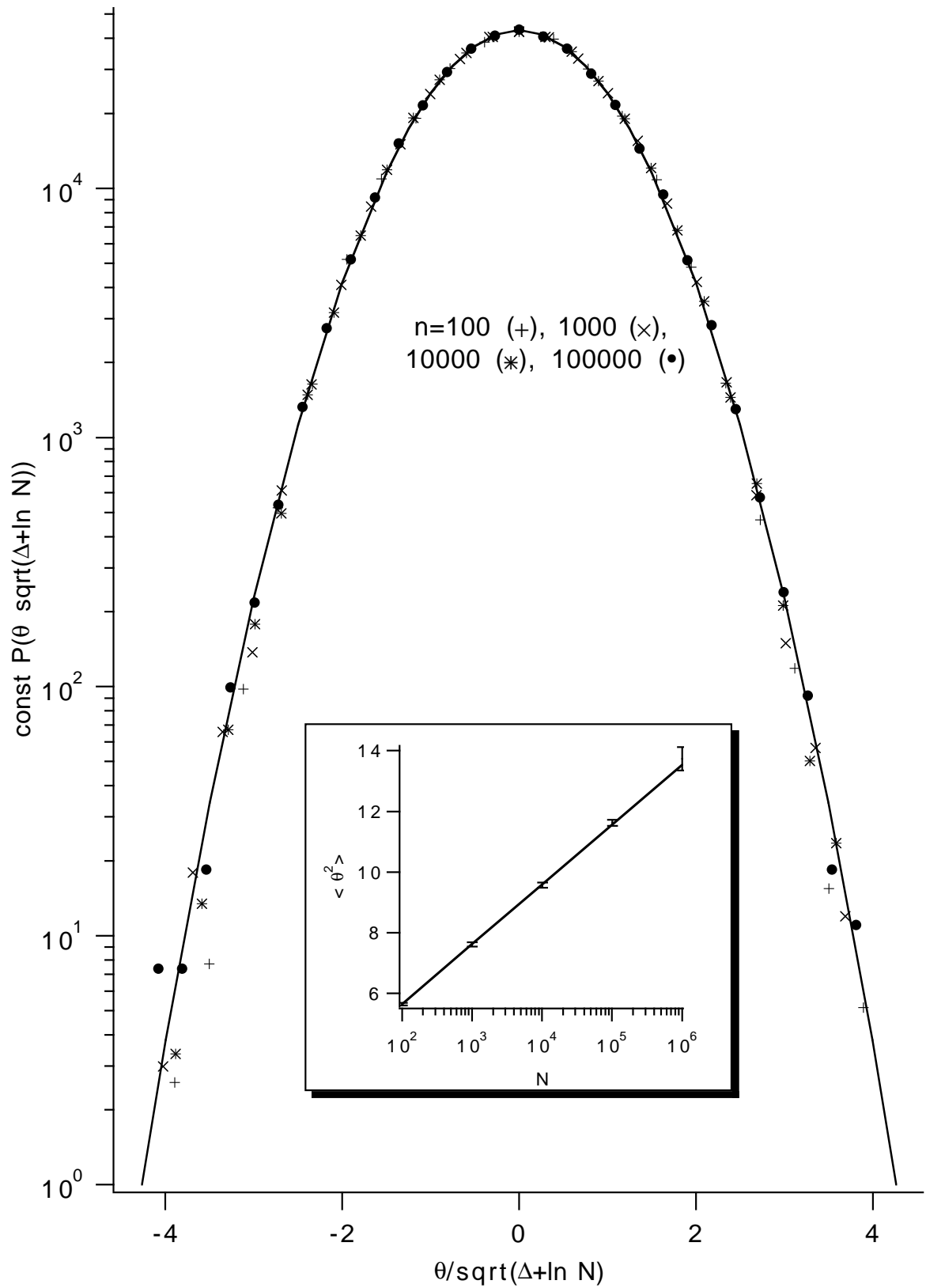,height=8.5in,angle=0}}
\caption{Winding angle distribution at the collapse transition on the Manhattan lattice}
\label{gauss}
\end{figure}

Our numerical simulations confirm nicely this result. Fig.~\ref{gauss} shows 
the winding angle distribution for polymers of length up to $10^5$, and 
the variance of the winding angle for length up to $10^6$. The solid line is the analytical result. 

In the swollen phase, we believe that the constant $C$ is $C=2$ for a SAW on the Manhattan lattice, i.e., the winding angle distribution is the same as for the normal SAW. The above calculation cannot be repeated in the swollen phase, since $\beta < 1$ in Eq.~(\ref{Z}) above. Although there exist still two paths connecting $i$ and $j$, they now have different weight. A path of $N$ steps will close to a loop after a mean number of steps that diverges, and therefore the second path has in fact no weight at all. We expect to arrive at the same situation as for a normal SAW, which is discussed in \cite{dup88}. 
Conformal field theory \cite{car94} suggests also that $C$ does not depend on the strength of the parallel interactions. Only the magnetic charges in Eq.~(\ref{G}) are affected by it, and these drop out when the winding angle is calculated. 

The calculation of this section can easily be generalized to ``watermelon configurations'', where the points $i$ and $j$ are connected by $L$ paths. The constant $C$ characterizing the winding angle distribution for one of these paths is then $C_L=4\nu/L^2g$. 

\section{Winding angles for collapsed polymers, and the phase diagram}
\label{arguments}

The above result $C=6/7$ was obtained in a situation where the polymer cannot make any parallel contacts. We therefore suggest that this corresponds to the case $\omega_p=0, \omega_a=\omega_\theta$. However, a SAW on a square lattice that has no parallel contacts is different from a SAW on a Mahattan lattice, since it can get trapped without loop formation. Only if the range of the repulsive parallel interaction is extended to next--nearest neighbors, the way back to the origin remains always open. If one assumes that all these situations are equivalent to each other, one must draw the conclusion that the precise range and form of the repulsive interaction between parallel bonds is not important, and 
that $C = 6/7$ holds on a finite part of the transition line. Assuming only one nonanalyticity on this line, we conclude that $C=6/7$ on the entire line $\omega_p<\omega_a$, and that it jumps to $C=24/7$ at the $\theta$--point. 
The $\theta$--point is consequently the point where the three phase boundaries meet. 

Of course, other interpretations are in principle possible, e.g., that the interacting SAW on the Manhattan lattice falls into a separate universality class. Our argument, however, is supported by our exact enumeration data. Although they do not allow a good estimate of the value of $C$, they show clearly that for both cases $\omega_p=1$ and $\omega_p=0$ the winding angle decreases during the collapse (i.e., with increasing $\omega_a$), while it increases during the collapse along the line $\omega_p=\omega_a$. The constant $C$ should therefore be smaller than 2 for $\omega_p < \omega_a = \omega_\theta$, and assuming universality, the conclusion $C=6/7$ follows naturally. It is the equivalent of the transfer matrix result that the exponent $\gamma$ assumes its Manhattan lattice value 6/7 for all $\omega_p < \omega_a = \omega_\theta$, and that it jumps to 8/7 at the $\theta$--point $\omega_p = \omega_a = \omega_\theta$. 

Our universality conjecture finds its strongest support when one studies the collapsed phase of the polymer. In the following, we argue that $C=0$ in the collapsed phase whenever $\omega_p < \omega_a$ ($C=0$ should also hold for a collapsed polymer on the Manhattan lattice), and $C=\infty$ when $\omega_p > \omega_a$. We start with 
the assumption that the winding angle distribution does not change when one goes slightly to the right or to the left of the line $\omega_p=\omega_a$, and we lead this assumption to a contradiction. 

On the diagonal $\omega_p=\omega_a$, the winding angle distribution is Gaussian with some unknown but finite constant $C$. Since $C$ increases when going from the swollen phase to the $\theta$--point, it certainly becomes even larger in the low--temperature phase. A polymer in the collapsed phase on the diagonal $\omega_p=\omega_a$ has a finite density of contacts along its trajectory, leading to $d m_p \propto R  d\theta$, and (with $d\theta \propto d\ln N$ and $R \propto N^{1/2}$) to $m_p \propto N^{1/2}$.  

\begin{figure}
\centerline{\psfig{file=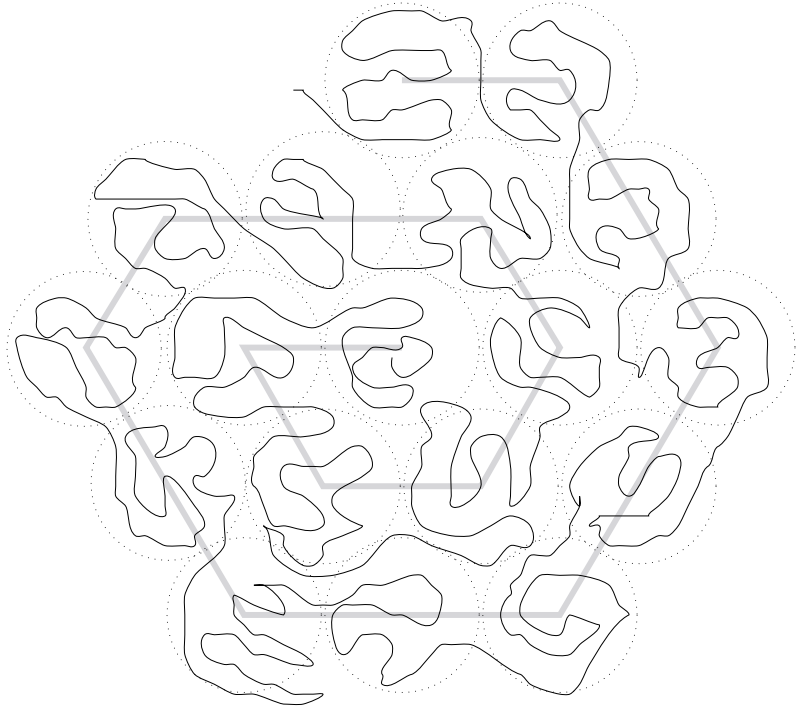,height=3.7in,angle=-90}}
\caption{A spiral composed of globules}
\label{ball}
\end{figure}

Let us first consider the case $\omega_p > \omega_a$, with  $\omega_p-\omega_a=\Delta \omega \ll 1$. We compare the free energy of a collapsed polymer in a globule configuration similar to the one on the line  $\omega_p=\omega_a$ to the free energy
of a spiral that is composed of globules of $n$ monomers (see Fig.~\ref{ball}). 
The difference in internal energy between the two is (neglecting constant coefficients)
$$\Delta U \simeq T\Delta \omega\left[\sqrt{N} - (N/n) \sqrt{n}\right] \, . $$
When one transforms a globule to a spiral, one breaks $O(\sqrt{N})$ parallel contacts, and one creates $O(N\sqrt{n})$ new parallel contacts. 
The leading contribution to the difference in entropy is 
$$\Delta S \simeq - (N/n) \ln n\,. $$
This is the number of globules times the entropy loss per globule when the end points of the polymer are restricted to the surface of the globule (see Sec.~\ref{partition}). The entropy difference between one large globule  of $N$ monomers and $N/n$ globules of $n$ monomers (without any constraint for the end points) increases slower than $O(N)$, since the entropy is an extensive quantity.
For any $\Delta \omega$, the gain in binding energy $-U$ is larger than the loss in entropy, when $n$ is sufficiently large. By minimizing
the free energy $\Delta F = \Delta U - T \Delta S$, we find (to leading order)
\begin{equation}
n \propto \Delta \omega^{-2}\, . \label{globulesize}
\end{equation}
The correlation length $\xi\propto \sqrt{n}$, which is proportional to the globule radius, diverges as $\xi \propto (\Delta \omega)^{-4}$. These results are correct for $N \gg n \gg 1$.

Spirals have a considerable entropy close to the transition. However, it is difficult to see the continuous character of this phase transition in simulations. The spiral shape can only be seen when polymer length is much larger than the globule size $n$. The transition appears to be shifted to the right by a distance $\Delta \omega  = O(1/\sqrt{N})$, as in \cite{ben95}.

Now, we consider the case  $\omega_p < \omega_a$, with  $\omega_a-\omega_p=\Delta \omega \ll 1$. We compare the free energy of a collapsed polymer in a configuration similar to the one on the line  $\omega_p=\omega_a$ to the free energy
of a polymer that has its end points at the surface. Bringing the end points at the surface, replaces $O(\sqrt{N})$ parallel contacts by antiparallel contacts, decreasing the internal energy by an amount proportional to $\Delta U \simeq \Delta \omega \sqrt{N}$. The entropy loss due to this restriction is of the order $\ln N$, as shown previously. Having the end points at the surface means always a decrease in free energy in the thermodynamic limit $N \to \infty$. This means that the winding angle and the number of parallel contacts must saturate in the thermodynamic limit. Only an initial and finite segment of length $n \propto (\Delta \omega)^{-2}$ of the polymer may behave like a polymer at the transition line. As for the spiral phase, the correlation length diverges as 
$\xi \propto (\Delta \omega)^{-4}$.

\begin{figure}
\centerline{\psfig{file=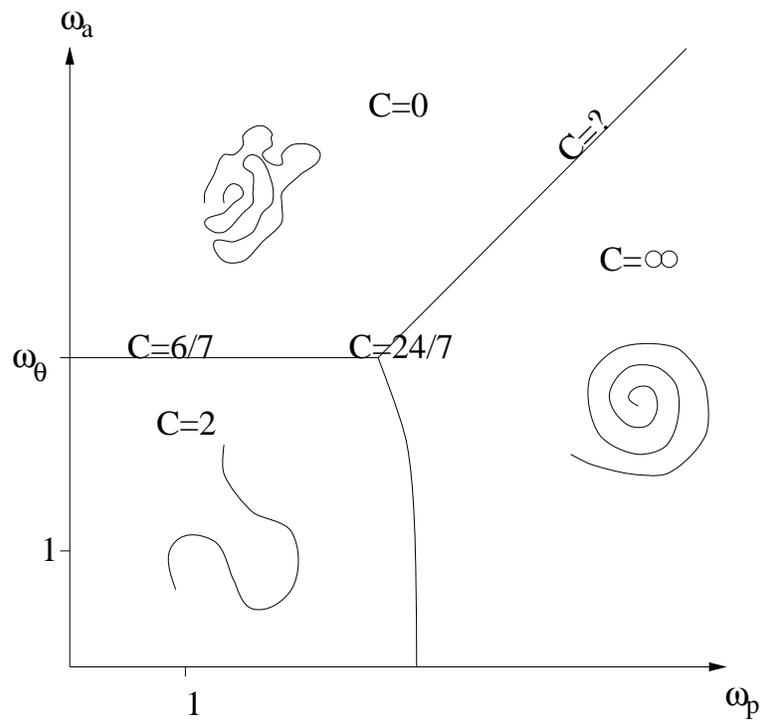,height=3.7in,angle=-90}}
\caption{Value of the constant C in most parts of the phase diagram}
\label{C}
\end{figure}

We thus have shown that the constant $C$ of the winding angle distribution in the collapsed phase has the same value $C=0$ for all $\omega_p < \omega_a$. It would be rather surprising if this universal feature did change at the collapse transition or in the swollen phase, where contacts play a less important role. The study of the collapsed phase therefore provides convincing support for the hypothesis that the winding angle distribution is a universal quantity. The constant $C$ characterizing the winding angle distribution can now be given for most of the phase diagram, as indicated in Fig.~\ref{C}. 

We have also seen that the number of parallel contacts saturates in the collapsed phase for $\omega_p < \omega_a$. 
This result should also hold in the swollen phase, where contacts are less important, leading to a universal exponent $\gamma$. 

\section{Summary and discussion}
\label{conclusion}

In this paper, we have studied oriented polymers with orientation--dependent interaction. We have argued that both the winding angle distribution and the exponent $\gamma$ are universal quantities, in agreement with transfer matrix calculations for the exponent $\gamma$ \cite{tro97}. This result is closely tied to the observation that the $\theta$--point is a special point in the phase diagram, where three phase boundaries meet. When parallel contacts have a different energy than antiparallel contacts, a symmetry is broken, and a phase transition takes place. In contrast to \cite{tro97}, we predict that this phase transition between the collapsed and the spiral phase is continuous. The existence of this phase transition is particular to two dimensions, since in higher dimensions a parallel contact can locally be transformed into an antiparallel contact, without changing the conformation of the polymer at a large scale. 

We  succeeded in obtaining the winding angle distribution for the case where antiparallel contacts dominate, at the collapse transition as well as in the low--temperature phase. 
We have also argued that the number of parallel contacts saturates in the thermodynamic limit whenever the antiparallel energy is larger than the parallel one. 

There are three challenges left: The winding angle distribution along the phase boundary between the collapsed and the spiral phase is still unknown. As mentioned in Sec.~\ref{windings}, collapsed polymers are different from dense polymers, for which $C=4$. 
Secondly, the crossover exponent $\Psi$ introduced in Eq.~(\ref{crossover}) needs to be determined.  Only if the $\theta$--point is a special point in the phase diagram, the number of parallel contacts increases with $N^\Psi$ at the $\theta$--point. Otherwise, it increases logarithmically in $N$, i.e., $\Psi$ vanishes. 
Finally, the transition from the swollen phase to the spiral phase is poorly understood. Although numerical results indicate a first--order transition \cite{fle95,bar96}, we cannot rule out a continuous transition. The assumption in \cite{fle95,bar96} that the spiral has no entropy cannot be upheld, as we have seen in our study of the globule--to--spiral transition. It is quite possible that the swollen--to--spiral transition has some analogy with the adsorption of a polymer at a wall, which is known to be continuous. The ``wall'' in our case is the surface of a spiral arm, and the thickness of the adsorbed polymer layer corresponds to the thickness of a spiral arm.

\acknowledgements
We thank J. Cardy for useful discussions, and A. Owczarek for alerting us to Ref.~\cite{tro97}. 
This work was supported by EPSRC Grant No.~GR/K79307.

\end{document}

%% file: psfig_19.tex
\def\PsfigVersion{1.9}
\ifx\undefined\psfig\else \fi

%

\let\LaTeXAtSign=\@
\let\@=\relax
\edef\psfigRestoreAt{\catcode`\@=\number\catcode`@\relax}
\catcode`\@=11\relax
\newwrite\@unused
\def\ps@typeout#1{{\let\protect\string\immediate\write\@unused{#1}}}
\ps@typeout{psfig/tex \PsfigVersion}


\def\figurepath{./}
\def\psfigurepath#1{\edef\figurepath{#1}}

%
%
\def\@nnil{\@nil}
\def\@empty{}
\def\@psdonoop#1\@@#2#3{}
\def\@psdo#1:=#2\do#3{\edef\@psdotmp{#2}\ifx\@psdotmp\@empty \else
    \expandafter\@psdoloop#2,\@nil,\@nil\@@#1{#3}\fi}
\def\@psdoloop#1,#2,#3\@@#4#5{\def#4{#1}\ifx #4\@nnil \else
       #5\def#4{#2}\ifx #4\@nnil \else#5\@ipsdoloop #3\@@#4{#5}\fi\fi}
\def\@ipsdoloop#1,#2\@@#3#4{\def#3{#1}\ifx #3\@nnil 
       \let\@nextwhile=\@psdonoop \else
      #4\relax\let\@nextwhile=\@ipsdoloop\fi\@nextwhile#2\@@#3{#4}}
\def\@tpsdo#1:=#2\do#3{\xdef\@psdotmp{#2}\ifx\@psdotmp\@empty \else
    \@tpsdoloop#2\@nil\@nil\@@#1{#3}\fi}
\def\@tpsdoloop#1#2\@@#3#4{\def#3{#1}\ifx #3\@nnil 
       \let\@nextwhile=\@psdonoop \else
      #4\relax\let\@nextwhile=\@tpsdoloop\fi\@nextwhile#2\@@#3{#4}}
%
\ifx\undefined\fbox
\newdimen\fboxrule
\newdimen\fboxsep
\newdimen\ps@tempdima
\newbox\ps@tempboxa
\fboxsep = 3pt
\fboxrule = .4pt
\long\def\fbox#1{\leavevmode\setbox\ps@tempboxa\hbox{#1}\ps@tempdima\fboxrule
    \advance\ps@tempdima \fboxsep \advance\ps@tempdima \dp\ps@tempboxa
   \hbox{\lower \ps@tempdima\hbox
  {\vbox{\hrule height \fboxrule
          \hbox{\vrule width \fboxrule \hskip\fboxsep
          \vbox{\vskip\fboxsep \box\ps@tempboxa\vskip\fboxsep}\hskip 
                 \fboxsep\vrule width \fboxrule}
                 \hrule height \fboxrule}}}}
\fi
%
%
\newread\ps@stream
\newif\ifnot@eof       
\newif\if@noisy        
\newif\if@atend        
\newif\if@psfile       
%
%
{\catcode`\%=12\global\gdef\epsf@start{
\def\epsf@PS{PS}
\def\epsf@getbb#1{%
%
%
\openin\ps@stream=#1
\ifeof\ps@stream\ps@typeout{Error, File #1 not found}\else
%
%
   {\not@eoftrue \chardef\other=12
    \def\do##1{\catcode`##1=\other}\dospecials \catcode`\ =10
    \loop
       \if@psfile
	  \read\ps@stream to \epsf@fileline
       \else{
	  \obeyspaces
          \read\ps@stream to \epsf@tmp\global\let\epsf@fileline\epsf@tmp}
       \fi
       \ifeof\ps@stream\not@eoffalse\else
%
%
       \if@psfile\else
       \expandafter\epsf@test\epsf@fileline:. \\%
       \fi
%
%
          \expandafter\epsf@aux\epsf@fileline:. \\%
       \fi
   \ifnot@eof\repeat
   }\closein\ps@stream\fi}%
%
%
\long\def\epsf@test#1#2#3:#4\\{\def\epsf@testit{#1#2}
			\ifx\epsf@testit\epsf@start\else
\ps@typeout{Warning! File does not start with `\epsf@start'.  It may not be a PostScript file.}
			\fi
			\@psfiletrue} 
%
%
{\catcode`\%=12\global\let\epsf@percent=
%
%
%
\long\def\epsf@aux#1#2:#3\\{\ifx#1\epsf@percent
   \def\epsf@testit{#2}\ifx\epsf@testit\epsf@bblit
	\@atendfalse
        \epsf@atend #3 . \\%
	\if@atend	
	   \if@verbose{
		\ps@typeout{psfig: found `(atend)'; continuing search}
	   }\fi
        \else
        \epsf@grab #3 . . . \\%
        \not@eoffalse
        \global\no@bbfalse
        \fi
   \fi\fi}%
%
%
\def\epsf@grab #1 #2 #3 #4 #5\\{%
   \global\def\epsf@llx{#1}\ifx\epsf@llx\empty
      \epsf@grab #2 #3 #4 #5 .\\\else
   \global\def\epsf@lly{#2}%
   \global\def\epsf@urx{#3}\global\def\epsf@ury{#4}\fi}%
%
%
\def\epsf@atendlit{(atend)} 
\def\epsf@atend #1 #2 #3\\{%
   \def\epsf@tmp{#1}\ifx\epsf@tmp\empty
      \epsf@atend #2 #3 .\\\else
   \ifx\epsf@tmp\epsf@atendlit\@atendtrue\fi\fi}


\chardef\psletter = 11 
\chardef\other = 12

\newif \ifdebug 
\newif\ifc@mpute 
\c@mputetrue 

\let\then = \relax
\def\r@dian{pt }
\let\r@dians = \r@dian
\let\dimensionless@nit = \r@dian
\let\dimensionless@nits = \dimensionless@nit
\def\internal@nit{sp }
\let\internal@nits = \internal@nit
\newif\ifstillc@nverging
\def \Mess@ge #1{\ifdebug \then \message {#1} \fi}

{ 
	\catcode `\@ = \psletter
	\gdef \nodimen {\expandafter \n@dimen \the \dimen}
	\gdef \term #1 #2 #3%
	       {\edef \t@ {\the #1}
		\edef \t@@ {\expandafter \n@dimen \the #2\r@dian}%
		\t@rm {\t@} {\t@@} {#3}%
	       }
	\gdef \t@rm #1 #2 #3%
	       {{%
		\count 0 = 0
		\dimen 0 = 1 \dimensionless@nit
		\dimen 2 = #2\relax
		\Mess@ge {Calculating term #1 of \nodimen 2}%
		\loop
		\ifnum	\count 0 < #1
		\then	\advance \count 0 by 1
			\Mess@ge {Iteration \the \count 0 \space}%
			\Multiply \dimen 0 by {\dimen 2}%
			\Mess@ge {After multiplication, term = \nodimen 0}%
			\Divide \dimen 0 by {\count 0}%
			\Mess@ge {After division, term = \nodimen 0}%
		\repeat
		\Mess@ge {Final value for term #1 of 
				\nodimen 2 \space is \nodimen 0}%
		\xdef \Term {#3 = \nodimen 0 \r@dians}%
		\aftergroup \Term
	       }}
	\catcode `\p = \other
	\catcode `\t = \other
	\gdef \n@dimen #1pt{#1} 
}

\def \Divide #1by #2{\divide #1 by #2} 

\def \Multiply #1by #2
       {{
	\count 0 = #1\relax
	\count 2 = #2\relax
	\count 4 = 65536
	\Mess@ge {Before scaling, count 0 = \the \count 0 \space and
			count 2 = \the \count 2}%
	\ifnum	\count 0 > 32767 
	\then	\divide \count 0 by 4
		\divide \count 4 by 4
	\else	\ifnum	\count 0 < -32767
		\then	\divide \count 0 by 4
			\divide \count 4 by 4
		\else
		\fi
	\fi
	\ifnum	\count 2 > 32767 
	\then	\divide \count 2 by 4
		\divide \count 4 by 4
	\else	\ifnum	\count 2 < -32767
		\then	\divide \count 2 by 4
			\divide \count 4 by 4
		\else
		\fi
	\fi
	\multiply \count 0 by \count 2
	\divide \count 0 by \count 4
	\xdef \product {#1 = \the \count 0 \internal@nits}%
	\aftergroup \product
       }}

\def\r@duce{\ifdim\dimen0 > 90\r@dian \then   
		\multiply\dimen0 by -1
		\advance\dimen0 by 180\r@dian
		\r@duce
	    \else \ifdim\dimen0 < -90\r@dian \then  
		\advance\dimen0 by 360\r@dian
		\r@duce
		\fi
	    \fi}

\def\Sine#1%
       {{%
	\dimen 0 = #1 \r@dian
	\r@duce
	\ifdim\dimen0 = -90\r@dian \then
	   \dimen4 = -1\r@dian
	   \c@mputefalse
	\fi
	\ifdim\dimen0 = 90\r@dian \then
	   \dimen4 = 1\r@dian
	   \c@mputefalse
	\fi
	\ifdim\dimen0 = 0\r@dian \then
	   \dimen4 = 0\r@dian
	   \c@mputefalse
	\fi
	\ifc@mpute \then
		\divide\dimen0 by 180
		\dimen0=3.141592654\dimen0
		\dimen 2 = 3.1415926535897963\r@dian 
		\divide\dimen 2 by 2 
		\Mess@ge {Sin: calculating Sin of \nodimen 0}%
		\count 0 = 1 
		\dimen 2 = 1 \r@dian 
		\dimen 4 = 0 \r@dian 
		\loop
			\ifnum	\dimen 2 = 0 
			\then	\stillc@nvergingfalse 
			\else	\stillc@nvergingtrue
			\fi
			\ifstillc@nverging 
			\then	\term {\count 0} {\dimen 0} {\dimen 2}%
				\advance \count 0 by 2
				\count 2 = \count 0
				\divide \count 2 by 2
				\ifodd	\count 2 
				\then	\advance \dimen 4 by \dimen 2
				\else	\advance \dimen 4 by -\dimen 2
				\fi
		\repeat
	\fi		
			\xdef \sine {\nodimen 4}%
       }}

\def\Cosine#1{\ifx\sine\UnDefined\edef\Savesine{\relax}\else
		             \edef\Savesine{\sine}\fi
	{\dimen0=#1\r@dian\advance\dimen0 by 90\r@dian
	 \Sine{\nodimen 0}
	 \xdef\cosine{\sine}
	 \xdef\sine{\Savesine}}}	      

\def\psdraft{
	\def\@psdraft{0}
}
\def\psfull{
	\def\@psdraft{100}
}

\psfull

\newif\if@scalefirst
\def\psscalefirst{\@scalefirsttrue}
\def\psrotatefirst{\@scalefirstfalse}
\psrotatefirst

\newif\if@draftbox
\def\psnodraftbox{
	\@draftboxfalse
}
\def\psdraftbox{
	\@draftboxtrue
}
\@draftboxtrue

\newif\if@prologfile
\newif\if@postlogfile
\def\pssilent{
	\@noisyfalse
}
\def\psnoisy{
	\@noisytrue
}
\psnoisy
\newif\if@bbllx
\newif\if@bblly
\newif\if@bburx
\newif\if@bbury
\newif\if@height
\newif\if@width
\newif\if@rheight
\newif\if@rwidth
\newif\if@angle
\newif\if@clip
\newif\if@verbose
\def\@p@@sclip#1{\@cliptrue}

\newif\if@decmpr


\def\@p@@sfigure#1{\def\@p@sfile{null}\def\@p@sbbfile{null}
	        \openin1=#1.bb
		\ifeof1\closein1
	        	\openin1=\figurepath#1.bb
			\ifeof1\closein1
			        \openin1=#1
				\ifeof1\closein1%
				       \openin1=\figurepath#1
					\ifeof1
					   \ps@typeout{Error, File #1 not found}
						\if@bbllx\if@bblly
				   		\if@bburx\if@bbury
			      				\def\@p@sfile{#1}%
			      				\def\@p@sbbfile{#1}%
							\@decmprfalse
				  	   	\fi\fi\fi\fi
					\else\closein1
				    		\def\@p@sfile{\figurepath#1}%
				    		\def\@p@sbbfile{\figurepath#1}%
						\@decmprfalse
	                       		\fi%
			 	\else\closein1%
					\def\@p@sfile{#1}
					\def\@p@sbbfile{#1}
					\@decmprfalse
			 	\fi
			\else
				\def\@p@sfile{\figurepath#1}
				\def\@p@sbbfile{\figurepath#1.bb}
				\@decmprtrue
			\fi
		\else
			\def\@p@sfile{#1}
			\def\@p@sbbfile{#1.bb}
			\@decmprtrue
		\fi}

\def\@p@@sfile#1{\@p@@sfigure{#1}}

\def\@p@@sbbllx#1{
		\@bbllxtrue
		\dimen100=#1
		\edef\@p@sbbllx{\number\dimen100}
}
\def\@p@@sbblly#1{
		\@bbllytrue
		\dimen100=#1
		\edef\@p@sbblly{\number\dimen100}
}
\def\@p@@sbburx#1{
		\@bburxtrue
		\dimen100=#1
		\edef\@p@sbburx{\number\dimen100}
}
\def\@p@@sbbury#1{
		\@bburytrue
		\dimen100=#1
		\edef\@p@sbbury{\number\dimen100}
}
\def\@p@@sheight#1{
		\@heighttrue
		\dimen100=#1
   		\edef\@p@sheight{\number\dimen100}
}
\def\@p@@swidth#1{
		\@widthtrue
		\dimen100=#1
		\edef\@p@swidth{\number\dimen100}
}
\def\@p@@srheight#1{
		\@rheighttrue
		\dimen100=#1
		\edef\@p@srheight{\number\dimen100}
}
\def\@p@@srwidth#1{
		\@rwidthtrue
		\dimen100=#1
		\edef\@p@srwidth{\number\dimen100}
}
\def\@p@@sangle#1{
		\@angletrue
		\edef\@p@sangle{#1} 
}
\def\@p@@ssilent#1{ 
		\@verbosefalse
}
\def\@p@@sprolog#1{\@prologfiletrue\def\@prologfileval{#1}}
\def\@p@@spostlog#1{\@postlogfiletrue\def\@postlogfileval{#1}}
\def\@cs@name#1{\csname #1\endcsname}
\def\@setparms#1=#2,{\@cs@name{@p@@s#1}{#2}}
%
%
\def\ps@init@parms{
		\@bbllxfalse \@bbllyfalse
		\@bburxfalse \@bburyfalse
		\@heightfalse \@widthfalse
		\@rheightfalse \@rwidthfalse
		\def\@p@sbbllx{}\def\@p@sbblly{}
		\def\@p@sbburx{}\def\@p@sbbury{}
		\def\@p@sheight{}\def\@p@swidth{}
		\def\@p@srheight{}\def\@p@srwidth{}
		\def\@p@sangle{0}
		\def\@p@sfile{} \def\@p@sbbfile{}
		\def\@p@scost{10}
		\def\@sc{}
		\@prologfilefalse
		\@postlogfilefalse
		\@clipfalse
		\if@noisy
			\@verbosetrue
		\else
			\@verbosefalse
		\fi
}
%
%
\def\parse@ps@parms#1{
	 	\@psdo\@psfiga:=#1\do
		   {\expandafter\@setparms\@psfiga,}}
%
%
\newif\ifno@bb
\def\bb@missing{
	\if@verbose{
		\ps@typeout{psfig: searching \@p@sbbfile \space  for bounding box}
	}\fi
	\no@bbtrue
	\epsf@getbb{\@p@sbbfile}
        \ifno@bb \else \bb@cull\epsf@llx\epsf@lly\epsf@urx\epsf@ury\fi
}	
\def\bb@cull#1#2#3#4{
	\dimen100=#1 bp\edef\@p@sbbllx{\number\dimen100}
	\dimen100=#2 bp\edef\@p@sbblly{\number\dimen100}
	\dimen100=#3 bp\edef\@p@sbburx{\number\dimen100}
	\dimen100=#4 bp\edef\@p@sbbury{\number\dimen100}
	\no@bbfalse
}
\newdimen\p@intvaluex
\newdimen\p@intvaluey
\def\rotate@#1#2{{\dimen0=#1 sp\dimen1=#2 sp
		  \global\p@intvaluex=\cosine\dimen0
		  \dimen3=\sine\dimen1
		  \global\advance\p@intvaluex by -\dimen3
		  \global\p@intvaluey=\sine\dimen0
		  \dimen3=\cosine\dimen1
		  \global\advance\p@intvaluey by \dimen3
		  }}
\def\compute@bb{
		\no@bbfalse
		\if@bbllx \else \no@bbtrue \fi
		\if@bblly \else \no@bbtrue \fi
		\if@bburx \else \no@bbtrue \fi
		\if@bbury \else \no@bbtrue \fi
		\ifno@bb \bb@missing \fi
		\ifno@bb \ps@typeout{FATAL ERROR: no bb supplied or found}
			\no-bb-error
		\fi
		%
%
		\count203=\@p@sbburx
		\count204=\@p@sbbury
		\advance\count203 by -\@p@sbbllx
		\advance\count204 by -\@p@sbblly
		\edef\ps@bbw{\number\count203}
		\edef\ps@bbh{\number\count204}
		\if@angle 
			\Sine{\@p@sangle}\Cosine{\@p@sangle}
	        	{\dimen100=\maxdimen\xdef\r@p@sbbllx{\number\dimen100}
					    \xdef\r@p@sbblly{\number\dimen100}
			                    \xdef\r@p@sbburx{-\number\dimen100}
					    \xdef\r@p@sbbury{-\number\dimen100}}
%
                        \def\minmaxtest{
			   \ifnum\number\p@intvaluex<\r@p@sbbllx
			      \xdef\r@p@sbbllx{\number\p@intvaluex}\fi
			   \ifnum\number\p@intvaluex>\r@p@sbburx
			      \xdef\r@p@sbburx{\number\p@intvaluex}\fi
			   \ifnum\number\p@intvaluey<\r@p@sbblly
			      \xdef\r@p@sbblly{\number\p@intvaluey}\fi
			   \ifnum\number\p@intvaluey>\r@p@sbbury
			      \xdef\r@p@sbbury{\number\p@intvaluey}\fi
			   }
			\rotate@{\@p@sbbllx}{\@p@sbblly}
			\minmaxtest
			\rotate@{\@p@sbbllx}{\@p@sbbury}
			\minmaxtest
			\rotate@{\@p@sbburx}{\@p@sbblly}
			\minmaxtest
			\rotate@{\@p@sbburx}{\@p@sbbury}
			\minmaxtest
			\edef\@p@sbbllx{\r@p@sbbllx}\edef\@p@sbblly{\r@p@sbblly}
			\edef\@p@sbburx{\r@p@sbburx}\edef\@p@sbbury{\r@p@sbbury}
		\fi
		\count203=\@p@sbburx
		\count204=\@p@sbbury
		\advance\count203 by -\@p@sbbllx
		\advance\count204 by -\@p@sbblly
		\edef\@bbw{\number\count203}
		\edef\@bbh{\number\count204}
}
%
%
\def\in@hundreds#1#2#3{\count240=#2 \count241=#3
		     \count100=\count240	
		     \divide\count100 by \count241
		     \count101=\count100
		     \multiply\count101 by \count241
		     \advance\count240 by -\count101
		     \multiply\count240 by 10
		     \count101=\count240	
		     \divide\count101 by \count241
		     \count102=\count101
		     \multiply\count102 by \count241
		     \advance\count240 by -\count102
		     \multiply\count240 by 10
		     \count102=\count240	
		     \divide\count102 by \count241
		     \count200=#1\count205=0
		     \count201=\count200
			\multiply\count201 by \count100
		 	\advance\count205 by \count201
		     \count201=\count200
			\divide\count201 by 10
			\multiply\count201 by \count101
			\advance\count205 by \count201
		     \count201=\count200
			\divide\count201 by 100
			\multiply\count201 by \count102
			\advance\count205 by \count201
		     \edef\@result{\number\count205}
}
\def\compute@wfromh{
		\in@hundreds{\@p@sheight}{\@bbw}{\@bbh}
		\edef\@p@swidth{\@result}
}
\def\compute@hfromw{
	        \in@hundreds{\@p@swidth}{\@bbh}{\@bbw}
		\edef\@p@sheight{\@result}
}
\def\compute@handw{
		\if@height 
			\if@width
			\else
				\compute@wfromh
			\fi
		\else 
			\if@width
				\compute@hfromw
			\else
				\edef\@p@sheight{\@bbh}
				\edef\@p@swidth{\@bbw}
			\fi
		\fi
}
\def\compute@resv{
		\if@rheight \else \edef\@p@srheight{\@p@sheight} \fi
		\if@rwidth \else \edef\@p@srwidth{\@p@swidth} \fi
}
%
\def\compute@sizes{
	\compute@bb
	\if@scalefirst\if@angle
	\if@width
	   \in@hundreds{\@p@swidth}{\@bbw}{\ps@bbw}
	   \edef\@p@swidth{\@result}
	\fi
	\if@height
	   \in@hundreds{\@p@sheight}{\@bbh}{\ps@bbh}
	   \edef\@p@sheight{\@result}
	\fi
	\fi\fi
	\compute@handw
	\compute@resv}

%
%
\def\psfig#1{\vbox {
	%
	\ps@init@parms
	\parse@ps@parms{#1}
	\compute@sizes
	\ifnum\@p@scost<\@psdraft{
		\special{ps::[begin] 	\@p@swidth \space \@p@sheight \space
				\@p@sbbllx \space \@p@sbblly \space
				\@p@sbburx \space \@p@sbbury \space
				startTexFig \space }
		\if@angle
			\special {ps:: \@p@sangle \space rotate \space} 
		\fi
		\if@clip{
			\if@verbose{
				\ps@typeout{(clip)}
			}\fi
			\special{ps:: doclip \space }
		}\fi
		\if@prologfile
		    \special{ps: plotfile \@prologfileval \space } \fi
		\if@decmpr{
			\if@verbose{
				\ps@typeout{psfig: including \@p@sfile.Z \space }
			}\fi
			\special{ps: plotfile "`zcat \@p@sfile.Z" \space }
		}\else{
			\if@verbose{
				\ps@typeout{psfig: including \@p@sfile \space }
			}\fi
			\special{ps: plotfile \@p@sfile \space }
		}\fi
		\if@postlogfile
		    \special{ps: plotfile \@postlogfileval \space } \fi
		\special{ps::[end] endTexFig \space }
		\vbox to \@p@srheight sp{
			\hbox to \@p@srwidth sp{
				\hss
			}
		\vss
		}
	}\else{
		\if@draftbox{		
			\hbox{\frame{\vbox to \@p@srheight sp{
			\vss
			\hbox to \@p@srwidth sp{ \hss \@p@sfile \hss }
			\vss
			}}}
		}\else{
			\vbox to \@p@srheight sp{
			\vss
			\hbox to \@p@srwidth sp{\hss}
			\vss
			}
		}\fi

	}\fi
}}
\psfigRestoreAt
\let\@=\LaTeXAtSign